\documentclass[aps,prb,reprint,amsmath,showpacs,superscriptaddress, longbibliography,nofootinbib]{revtex4-2}
\usepackage{graphicx}
\usepackage{textcomp}
\usepackage{float}
\usepackage{xcolor}
\usepackage[normalem]{ulem}
\usepackage{upgreek}

\newcommand{\vF}{v_{\mathrm{F}}}

\newcommand{\wG}{\omega_\mathrm{G}}
\newcommand{\wD}{\omega_\mathrm{2D}}
\newcommand{\AD}{A_\mathrm{2D}}
\newcommand{\WG}{\Gamma_\mathrm{G}}
\newcommand{\WD}{\Gamma_\mathrm{2D}}

\newcommand{\EF}{E_\mathrm{F}}

\newcommand{\cm}{\mathrm{cm^{-1}}}
\newcommand{\etal}{\textit{et~al.~}}     

\begin{document}

\title{Charge carrier density-dependent Raman spectra of graphene \\ encapsulated in hexagonal boron nitride}

\author{J.~Sonntag}
     \affiliation{JARA-FIT and 2nd Institute of Physics, RWTH Aachen University, 52074 Aachen, Germany}
    \affiliation{Peter Gr{\"u}nberg Institute (PGI-9), Forschungszentrum J{\"u}lich, 52425 J{\"u}lich, Germany}

\author{K.~Watanabe}
    \affiliation{Research Center for Functional Materials,
National Institute for Materials Science, 1-1 Namiki, Tsukuba 305-0044, Japan}

\author{T.~Taniguchi}
    \affiliation{International Center for Materials Nanoarchitectonics,
National Institute for Materials Science,  1-1 Namiki, Tsukuba 305-0044, Japan}

 \author{B. Beschoten}
     \affiliation{JARA-FIT and 2nd Institute of Physics, RWTH Aachen University, 52074 Aachen, Germany}
\affiliation{JARA-FIT Institute for Quantum Information, Forschungszentrum J\"ulich GmbH and RWTH Aachen University, 52074 Aachen, Germany}
\author{C.~Stampfer}
    \affiliation{JARA-FIT and 2nd Institute of Physics, RWTH Aachen University, 52074 Aachen, Germany}
    \affiliation{Peter Gr{\"u}nberg Institute (PGI-9), Forschungszentrum J{\"u}lich, 52425 J{\"u}lich, Germany}
    \email{Corresponding author: stampfer@physik.rwth-aachen.de}

\date{\today}

\begin{abstract}
	
We present low-temperature Raman measurements on gate tunable graphene encapsulated in hexagonal boron nitride,
which allows to study in detail
the Raman G and 2D mode frequencies and line widths as function of the charge carrier density. We observe a clear softening of the Raman G mode (of up to 2.5~cm$^{-1}$) at low carrier density due to the phonon anomaly and a residual G~mode line width of $\approx$~3.5~cm$^{-1}$ at high doping.
From analyzing the G mode dependence on doping and laser power we extract an electron-phonon-coupling constant of $\approx 4.4 \times10^{-3}$ (for the G mode phonon).
The ultra-flat nature of encapsulated graphene results in a minimum Raman 2D peak line width of 14.5~cm$^{-1}$ and allows to observe the intrinsic electron-electron scattering induced broadening of the 2D peak of up to 18~cm$^{-1}$ for an electron density of 5$\times$10$^{12}$~cm$^{-2}$ (laser excitation energy of 2.33~eV).
Our findings not only provide insights into electron-phonon coupling and the role of electron-electron scattering for the broadening of the 2D peak, but also crucially shows the limitations when it comes to the use of Raman spectroscopy (i.e. the use of the frequencies and the line widths of the G and 2D modes) to benchmark graphene in terms of charge carrier density, strain and strain inhomogenities.This is particularly relevant when utilizing spatially-resolved 2D Raman line width maps to assess substrate-induced nanometer-scale strain variations.

\end{abstract}

\maketitle

\begin{figure*}

    \includegraphics[draft=false,keepaspectratio=true,clip,width=2.0\columnwidth]{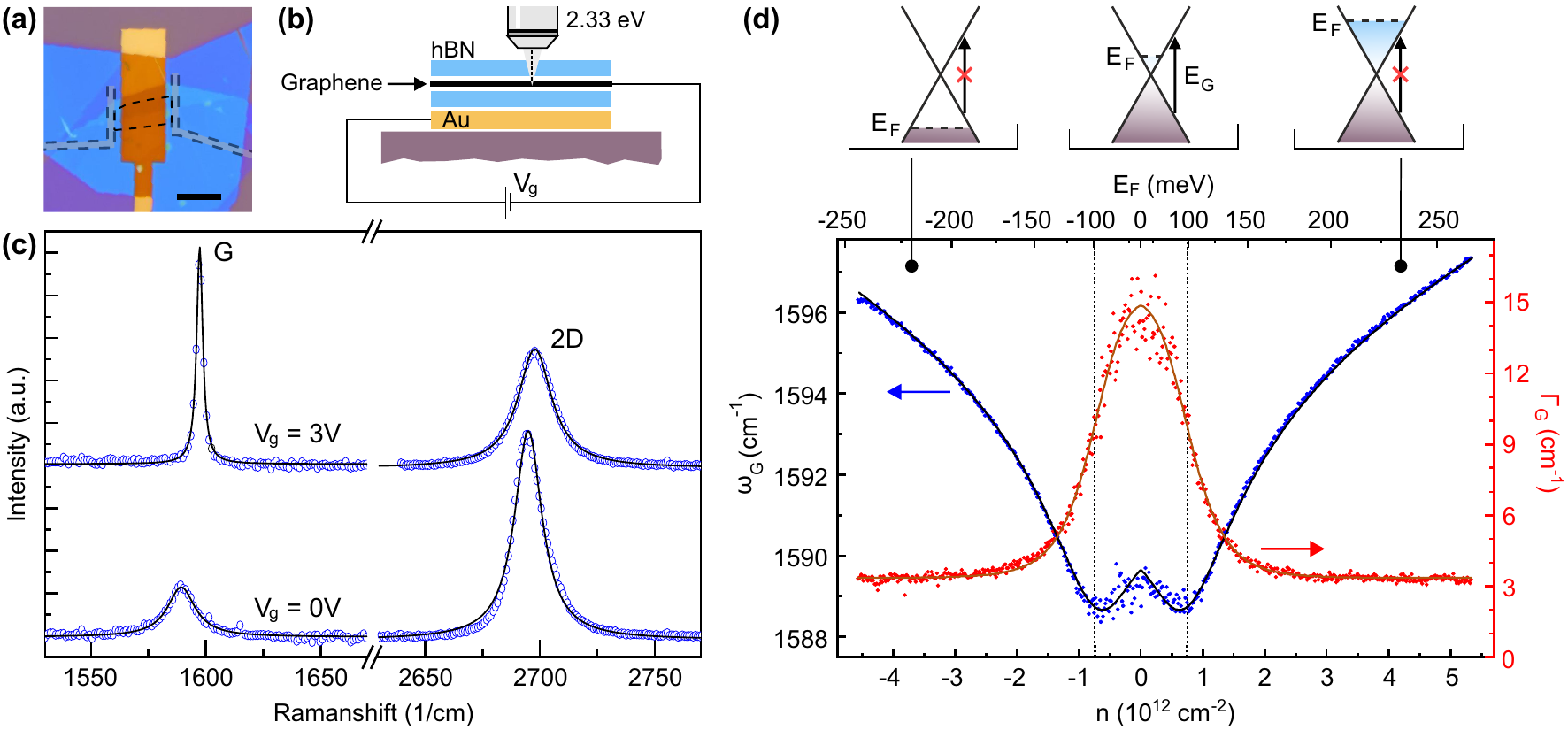}
	\caption{
		(a)	Optical image of the sample before contact fabrication. The black and grey outlines show the position of the graphene flake and contacts, respectively. The scale bar represents $10\,\upmu\mathrm{m}$.
		(b) Schematic illustration of the sample cross-section highlighting the hBN/graphene/hBN stack placed on top of the metal (Au) bottom gate, which allows to apply a gate voltage $V_g$.
		(c) The Raman spectrum of a high quality graphene/hBN heterostructure for two different gate voltages $V_g = 0$~V ($n\approx 0$, lower spectrum) and $V_g = 3$~V ($n\approx5.3 \times $10$^{12}$~cm$^{-2}$, upper spectrum).
		(d)
		Position, i.e. frequency $\wG$  (blue data points) and line width $\WG$ (red data points) of the Raman G peak as a function of the charge carrier density measured with a laser power of $1\,\mathrm{mW}$. The upper axis shows the corresponding Fermi energy $\EF$ under the assumption of $\vF = 0.98\times10^6\,\mathrm{m/s}$. The vertical dotted lines shows the resonance condition $|\EF|=E_\mathrm{G}/2=\hbar \vF\sqrt{\pi n}$. The solid lines represent fits based on Eq.~(\ref{eq:PhononAnomaly}) and Eq.~(\ref{eq:PhononAnomalyGamma}) to $\wG$ (black trace) and $\WG$ (brown trace), respectively, under the assumption of a finite effective temperature. The schematics on the top illustrate the Fermi energy dependent electron-phonon coupling for the G mode phonon.
	}
	\label{fig01}
\end{figure*}

\section{Introduction}

Raman spectroscopy is a highly useful method for the characterization and benchmarking of single- and few-layer graphene~\cite{Ferrari2006, Graf2007Feb,Cancado2011Aug,Lucchese2010Apr,Calizo2007Sep,Balandin2008Mar,Mohiuddin2009May,Lee2012,Mohr2009Nov,Huang2010,Frank2011,Tsoukleri2009Nov,Drogeler2014Nov,Neumann2015b,Banszerus2017} as well as graphene-based heterostructures, including twisted bilayer graphene~\cite{Carozo2011Nov,Gadelha2021Feb,Schapers2022Jul}. Among the parameters accessible via confocal Raman spectroscopy are the number of graphene layers~\cite{Ferrari2006, Graf2007Feb}, the amount of lattice defects~\cite{Cancado2011Aug,Lucchese2010Apr}, the lattice temperature~\cite{Calizo2007Sep,Balandin2008Mar}, the mechanical strain of the crystal structure~\cite{Mohiuddin2009May,Lee2012,Mohr2009Nov,Huang2010,Frank2011,Tsoukleri2009Nov,Drogeler2014Nov} and the amount of (substrate-induced) nanometer-scale strain variations~\cite{Neumann2015b,Banszerus2017}.
Thanks to the interrelation of the Raman spectra with the electronic structure of graphene, Raman spectroscopy also gives access to a variety of electronic parameters like the charge carrier density~\cite{Yan2007, Pisana2007,Stampfer2007,Froehlicher2015May,Das2008Mar,Lee2012}, defect densities and the expected maximum charge carrier mobility~\cite{Couto2014,Neumann2015b}.

It is therefore of high importance to understand the precise nature of the electron-phonon coupling in graphene and its evolution with charge carrier density. While this was already studied extensively for graphene on SiO\textsubscript{2}~\cite{Yan2007, Pisana2007,Stampfer2007,Froehlicher2015May,Das2008Mar,Lee2012} and partly for graphene on hexagonal boron nitride (hBN)~\cite{Chattrakun2013Nov}, a comprehensive investigation of the charge carrier density dependent Raman spectrum of high-quality, i.e. ultra-flat graphene with very low detrimental influence of the substrate, i.e. graphene encapsulated in hBN or suspended graphene, is still missing. So far, these measurements were prohibited by laser illumination-induced pinning of the Fermi energy to the charge neutrality point (CNP) in hBN/graphene/hBN heterostructures, because of so-called photodoping effects~\cite{Neumann2016Apr,Ju2014Apr}. These effects do not occur in suspended graphene. However, due to the electrostatic force induced by the electrostatic gate the Raman spectrum is in this case dominated by strain effects~\cite{Metten2016Dec} making a thorough investigation of doping effects unfeasible.

Here, we show how hBN/graphene/hBN heterostructures deposited on a local gold gate rather than on SiO\textsubscript{2} can be employed to thoroughly investigate the Raman spectra of state-of-the-art high quality, i.e. ultra-flat graphene as a function of charge carrier density. Thanks to the high quality of the hBN/graphene/hBN heterostructure and low temperatures, the Raman measurements presented in this work constitute
some of the most pristine Raman spectra of gated graphene measured so far.
In particular, we show in this work first the unambiguous experimental observation of the breakdown of the Born-Oppenheimer approximation and the consequent appearance of the phonon anomaly in the G-peak in graphene (section II).
Secondly, we focus on the 2D~peak (section III), which shows an ultimately narrow 2D line width of $\WD\approx14.5~\cm$. The pristineness nature of the 2D line shape allows us to observe the influence of electron-electron scattering not only on the intensity, but also on the width of the 2D~peak. This also allows to extract the phonon-coupling strength near the $K$ point.

Importantly, the insights on the 2D line width, specifically the electron-electron scattering-induced broadening, highlight a significant limitation in utilizing 2D line width Raman mapping for evaluating graphene samples and fabrication processes in terms of substrate-induced nanoscale strain variations~\cite{Neumann2015b,Banszerus2017}. This limitation is crucial to consider as strain variations negatively impact the charge carrier mobility in bulk graphene at low temperatures~\cite{Couto2014}.

\section{Anomaly of the G~mode phonon }

To investigate the charge carrier density-dependent Raman spectra of high-quality, i.e. ultra-flat graphene we use exfoliated hBN crystals for encapsulation. The resulting stack is deposited on a prefabricated gold (Au) bottom gate. Subsequently, we fabricate one-dimensional contacts to the hBN/graphene/hBN stack by reactive ion etching (RIE) and metal deposition (5~nm Cr/ 50~nm Au)~\cite{Wang2013}.
An optical microscope image of the resulting sample is shown in Fig.~1(a) and a cross-section of the plate capacitor-like sample structure, which allows to apply a gate voltage $V_g$ for controlling the charge carrier density, is illustrated in Fig.~1(b).
Importantly, this sample structure offers both high electronic quality and the possibility to tune the charge carrier density without any photodoping effects~\cite{Neumann2016Apr,Ju2014Apr} or gate-induced strain effects as seen in suspended graphene samples~\cite{Metten2016Dec}. For our measurement we utilize a commercial, confocal, low-temperature ($\sim$ 4.2~K) Raman setup.
We use linearly polarized laser light
(wavelength of $532\,\mathrm{nm}$) with a power of 1\,mW (if not stated otherwise) and a spot size of $\sim500\,\mathrm{nm}$.
The position on the sample was chosen so that the line width of the 2D peak, which reflects the amount of nanometer-scale strain variations~\cite{Neumann2015b}, is minimal and homogeneous over a range greater than $2 \times 2$~$\mathrm{\mu}$m$^2$.
The scattered light is detected by a CCD spectrometer with a grating of 1200\,lines/mm. In Fig.~\ref{fig01}(c) the Raman spectra around the G~peak and the 2D~peak is shown exemplary for low ($V_g=0$~V)
and high ($V_g=3$~V)
gate voltage, corresponding to a low and a high charge carrier density. Note that the gold gate induces a broad background signal, which we subtract by fitting a third-order polynomial background. Evidently, the G~peak narrows significantly and shifts to higher wave numbers with increasing doping. The 2D~peak decreases in intensity and shifts to higher wave numbers as well.
We point out that the 2D~peak shows a very low full width half maximum (FWHM) value of $\WD\approx14.5\,\cm$ near the charge neutrality point (CNP) (see also Fig.~3(a)), which illustrates the negligible amounts of strain variations and the high quality of our graphene sample~\cite{Couto2014,Neumann2015b}.
To the best of our knowledge, this is the lowest $\WD$ observed so far~\cite{Froehlicher2015May, Berciaud2013,Berciaud2009,Neumann2015b,Couto2014,Frank2011,Wang2017Oct,Chen2019Aug} and
it enables us to provide an unprecedented reference for the charge carrier density dependence of the Raman spectra of pristine graphene.
To this end we extract the position $\omega$ and line width $\Gamma$ as a function of the charge carrier density $n = \alpha (V_g - V_g^0)$ from Lorentzian fits for both the G and 2D~peak. Here, $\alpha = 1.8 \times 10^{12}$ 1/(V cm$^2$) is the gate lever arm determined from Landau fan measurements and $V_g^0 = 37$~mV describes the offset from the CNP also obtained from transport measurements (see Appendix A).
We ensure that our Lorentzian fits are not influenced by any residual background signal by fitting the sum of a Lorentzian and a linear function to the spectral area around the G and 2D~peak.

Figure~1(d) shows the extracted positions $\wG$ and line widths $\WG$ of the G~peak as a function of $n$. Note that the extracted data shown consists of multiple sweeps with increasing and decreasing $V_\mathrm{g}$.
As observed previously in numerous experimental studies~\cite{Yan2007, Pisana2007,Stampfer2007,Froehlicher2015May,Das2008Mar}, we find that the G~peak is heavily influenced by $n$. We observe the typical hallmarks of the resonant coupling of the G-mode phonon to the electronic transitions across the gapless bands of graphene located at half the phonon energy $|\EF|=E_\mathrm{G}/2=\hbar \vF \sqrt{\pi n}$, see illustrations in top row of Fig.~1(d). For large $|n|$, the phonon frequency $\wG$ increases due to the non-adiabatic phonon hardening~\cite{Lazzeri2006,Ando2006a} also observed previously~\cite{Froehlicher2015May,Pisana2007,Yan2007,Stampfer2007,Das2008Mar}.
Unlike the previous experimental studies, we unambiguously observe the predicted anomalous phonon softening at $|\EF|\approx E_\mathrm{G}/2$ (see dotted vertical lines in the main panel of Fig.~1(d)) due to the high electronic quality and homogeneity of our hBN/graphene/hBN sample.
It is noteworthy that the phonon softening gets significantly more pronounced when reducing the laser power (see Fig.~2(a)) reaching a dip of around 2.5~cm$^{-1}$ for a laser power $p$ of 0.01~mW. However, this softening
is still significantly weaker than the theoretically expected value of around 7.5~cm$^{-1}$ for $T=4.2\,\mathrm{K}$.

The electron-phonon coupling also allows the phonons to decay into electron-hole pairs, resulting in a limited lifetime and consequently in a high G-peak line width of $\WG\approx15\,\cm$ for low doping (see central top schematic in Fig.~1(d)).
For larger (or smaller) Fermi energies $|\EF|>E_\mathrm{G}/2$, these transitions are Pauli blocked (see left and right schematics in Fig.~1(d)) and the G-peak narrows to $\WG\approx3.5\,\cm$, which is one of the lowest values found experimentally and is approaching the broadening due to anharmonic contributions~\cite{Bonini2007Oct} consistent with the negligible amount of disorder-induced broadening in our sample.

The non-adiabatic frequency shift as a function of $n$ can be calculated as the real part of the self-energy (see~\cite{Lazzeri2006,Pisana2007,Ando2006a}) resulting in
\begin{equation}
	\hbar\Delta\wG = \lambda \mathrm{P}\int_{\infty}^{-\infty} \frac{ |f(E-E_\mathrm{F})-f(E)| E^2 sgn(E)}{E^2-(\hbar\omega_\mathrm{G}^0)^2/4}dE,
	\label{eq:PhononAnomaly}
\end{equation}
where $\lambda$ is the electron-phonon coupling constant, P being the Cauchy principal value, $\omega_\mathrm{G}^0=E_\mathrm{G}/\hbar$ is the frequency of the G~peak for pristine graphene and $f(E)$ is the Fermi-Dirac distribution. For $T=0\,\mathrm{K}$ the integral of Eq.~(\ref{eq:PhononAnomaly}) shows a logarithmic divergence when the Fermi energy matches exactly half the phonon energy $\EF=\pm\hbar\omega_\mathrm{G}^0/2\approx\pm 98\,\mathrm{meV}$ \cite{Lazzeri2006}. For elevated temperatures or high charge disorder the divergences, also called \textit{phonon anomalies}, are smeared out. The line width of the G~peak, $\WG$, can be similarly calculated as the imaginary part of the self-energy~\cite{Lazzeri2006,Pisana2007,Ando2006a} leading to
\begin{equation}
	\WG =  \frac{\pi\lambda\omega_\mathrm{G}^0 }{2}   \left[ f \left( - \frac{\hbar \omega_\mathrm{G}^0}{2}-E_\mathrm{F} \right) - f \left( \frac{\hbar \omega_\mathrm{G}^0}{2}-E_\mathrm{F} \right) \right] + \Gamma_\mathrm{G}^0,
	\label{eq:PhononAnomalyGamma}
\end{equation}
where $\Gamma_\mathrm{G}^0$ is the G~peak width resulting from all other broadening mechanisms, such as disorder, strain variations and anharmonic coupling.
Thus, the peak width shows a step like behavior with a high $\WG$ for all $|\EF|<\hbar\omega_\mathrm{G}^0/2$ and a sudden decrease at half the phonon energy. Temperature and charge disorder smear out these transitions, as indicated by the Fermi-Dirac distributions in Eq.~(\ref{eq:PhononAnomalyGamma}).
This again becomes experimentally most apparent for low laser powers (see Fig.~2(b)).

\begin{figure*}[t]
    \includegraphics[draft=false,keepaspectratio=true,clip,width=2.0\columnwidth]{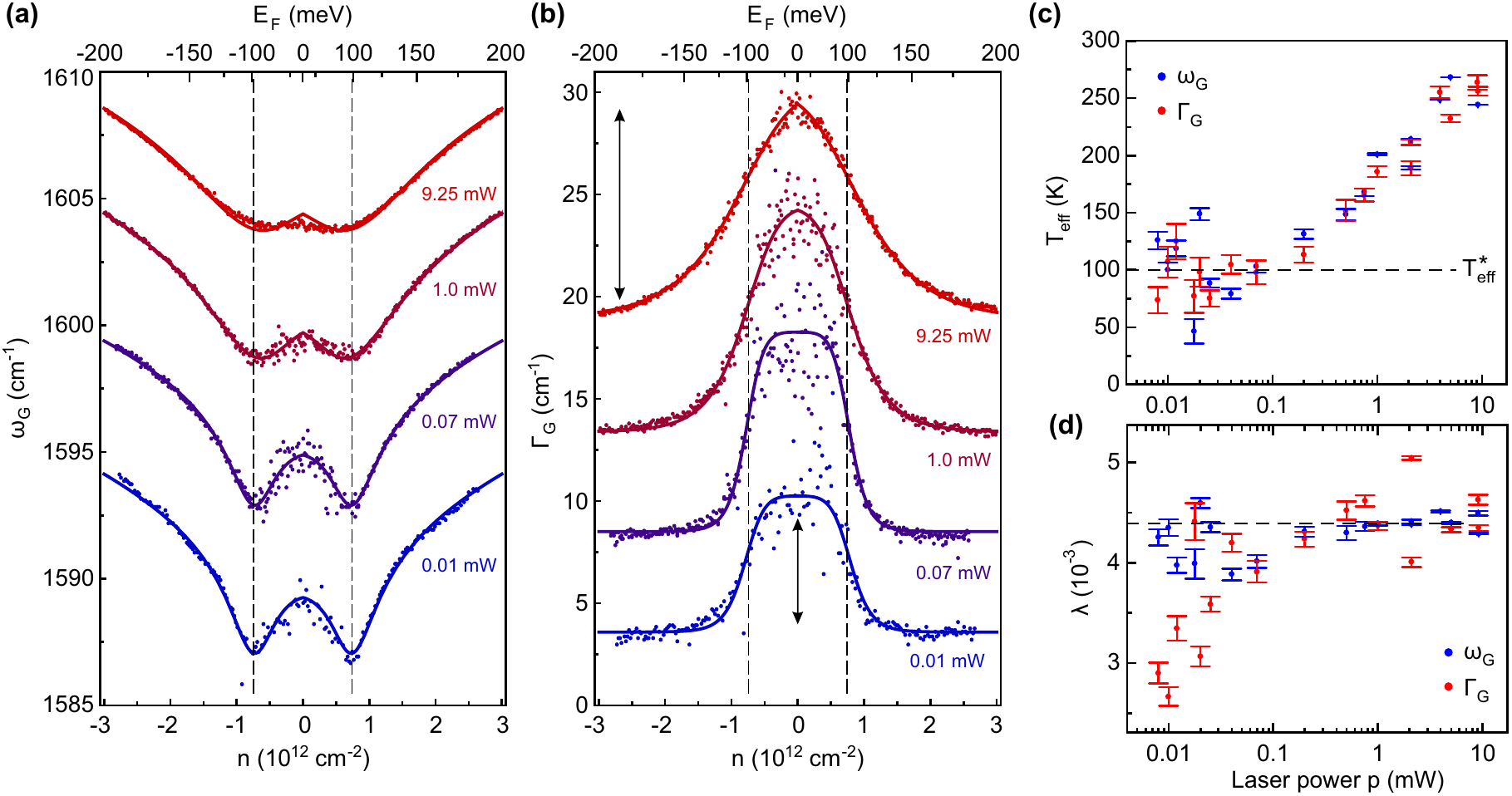}
	\caption{
	(a) G peak frequency $\wG$ as a function of charge carrier density $n$ for different magnitudes of laser power (see colored labels). The solid lines are fits based on Eq.~(\ref{eq:PhononAnomaly}). (b) Laser power dependence of $\WG$. Solid lines are fits based on Eq.~(\ref{eq:PhononAnomalyGamma}). For clarity each graph in panels (a) and (b) is offset by $5\,\cm$ and
	the vertical dashed lines shows the resonance condition $|\EF|=E_\mathrm{G}/2$ (as in Fig.~1(d)).
	(c) The extracted effective temperature $T_\mathrm{eff}$ as a function of laser power $p$ extracted from $\wG(n)$ (blue data points) and $\WG(n)$ (red data points) by fitting Eq.~(\ref{eq:PhononAnomaly}) and Eq.~(\ref{eq:PhononAnomalyGamma}), respectively. The dashed line marks the experimentally observed effective saturation temperature $T^*_\mathrm{eff}$.
	(d) Extracted electron-phonon-coupling constant $\lambda$ as a function of laser power; again from both $\wG(n)$ (blue data) and $\WG(n)$ (red data) obtained from fitting with Eq.~(\ref{eq:PhononAnomaly}) and Eq.~(\ref{eq:PhononAnomalyGamma}), respectively. Interestingly, $\lambda$ is constant when extracted from $\wG(n)$ and in good agreement with the value reported in Ref.~\cite{Pisana2007} (see dashed line), but decreases for low power when extracting it from $\WG(n)$.
	}
	\label{fig01}
\end{figure*}

Next we use Eqs.~(1) and (2) with the fitting parameters $T_\mathrm{eff}$, $\lambda$, $\vF$ and $\omega^0_\mathrm{G}$ or $\Gamma^0_\mathrm{G}$ to fit our experimental data $\wG(n)$ and $\WG(n)$, respectively.
For example, the black and brown solid trace in Fig.~1(d) show corresponding fits. Interestingly, from both fits we find an elevated effective temperature of $T_\mathrm{eff}=(201\pm0.6)\,\mathrm{K}$ from $\wG(n)$ (black trace) and $T_\mathrm{eff}=(185\pm5)\,\mathrm{K}$ from $\WG(n)$ (brown trace), owing to the relatively weak manifestation of the phonon anomalies at $|\EF|\approx E_\mathrm{G}/2$ and the smooth decrease of $\WG$ in our data.
Please note that here we use this elevated effective temperature to capture broadening of multiple origins, e.g. charge disorder within the laser spot, laser-induced heating of the electronic system, a finite lifetime and a related electronic broadening of the electronic states. Furthermore, we find the electron-phonon-coupling constant to be $\lambda=(4.38\pm0.01)\times10^{-3}$ from $\wG(n)$ and $\lambda=(4.37\pm0.04)\times10^{-3}$ from $\WG(n)$ in agreement with previous reports~\cite{Pisana2007,Froehlicher2015May}. The extracted Fermi velocities are $\vF=(0.944\pm0.02)\times10^6\,\mathrm{m/s}$ from $\wG(n)$ and $\vF=(0.930\pm0.04)\times10^6\,\mathrm{m/s}$ from $\WG(n)$.
Moreover we obtain $\wG^0=(1588.7\pm0.1)\,\cm$ and $\WG^0=(3.37\pm0.03)\,\cm$.

To elucidate the origin of the broadening expressed by the elevated effective temperature, especially with regard to laser-induced heating, we perform laser power dependent measurements.
In total we were able to vary the laser power $p$ by almost three orders of magnitude from $0.01\,\mathrm{mW}$ to $9.25\,\mathrm{mW}$, limited only by the required integration time for low laser powers. In Figs.~2(a) and~2(b) we show the position $\wG$ and the line width $\WG$ of the G-peak as a function of charge carrier density for different magnitudes of the laser power. With decreasing laser power, the decline in $\WG$ gets sharper and the phonon anomaly at $\EF=\hbar\wG/2$ becomes significantly more pronounced.
This reduced broadening for lower laser power indicates a decrease in the effective electron temperature $T_\mathrm{eff}$ due to a reduced effective laser-induced heating of the electronic system, as  seen in Fig.~2(c), where the extracted $T_\mathrm{eff}$ is plotted as function of laser power $p$; again from fitting with Eq.~(1) (blue data) and from fitting with Eq.~(2) (red data).
Interestingly, $T_\mathrm{eff}$ increases logarithmically above $p\approx 0.1~\mathrm{mW}$, whereas it saturates at $\sim100\,\mathrm{K}$ for lower laser powers.

Note that the independent fits to $\wG$ (Eq.~(1)) and $\WG$ (Eq.~(2)) give very similar results.
This shows that the electronic system can be significantly heated by the laser illumination and is in contrast to the lattice temperature which stays constant as indicated by the unchanging $\wG^{0}$ and $\WG^\mathrm{0}$~\cite{Calizo2007Sep,Balandin2008Mar}, see Fig.~6 in Appendix~B.
As the saturation value of $T^*_\mathrm{eff}\approx100\,\mathrm{K}$ at low power is significantly higher than the cryostat temperature of $4.2\,\mathrm{K}$ we attribute this discrepancy to the residual doping inhomogeneities in our sample.
Indeed, the effective temperature $T^*_\mathrm{eff}$ corresponds to an effective energy broadening of $\delta\EF=k_\mathrm{B}T^*_\mathrm{eff}\approx9\,\mathrm{meV}$ around the phonon anomaly, i.e. at $\EF\approx E_\mathrm{G}/2 = \hbar \vF \sqrt{\pi n}$. With the latter expression we estimate a corresponding charge carrier density disorder of $\delta n\approx10\times10^{10}\,\mathrm{cm^{-2}}$, which interestingly is in good agreement with the residual charge carrier density disorder $n^*\approx13\times10^{10}\,\mathrm{cm^{-2}}$ obtained from transport measurements on the very same sample after laser illumination (see Appendix A and Fig.~5).

Noteworthy, we not only find a decrease in $T_\mathrm{eff}$ for lower laser power $p$, but also a decrease in $\WG$ at the CNP, see black arrows in Fig.~2(b).
As described by Eq.~(\ref{eq:PhononAnomalyGamma}), this directly relates to a decreased electron-phonon-coupling constant $\lambda$ for low laser power $p$, as shown in Fig.~2(d) by the red data points. With increasing $p$ we see a $\sim50\,\%$ increase from $\lambda<3\times10^{-3}$ to a saturation value of $\lambda\approx4.4\times10^{-3}$ for $p>0.1\,\mathrm{mW}$. Surprisingly, this is in direct contrast to the electron-phonon-coupling constant as extracted from $\wG$, which does not show a dependence on the laser power (see blue data points in Fig.~2(d)) and remains at the saturation value of $\lambda\approx4.4\times 10^{-3}$ in good agreement with the value found previously~\cite{Pisana2007} (see dashed line in Fig.~2(d)).

The origin of this inconsistency in $\lambda$ between the two extraction methods  ($\wG$ (Eq.~(1)) and $\WG(n)$ (Eq.~(2))), might be due to the different sensitivity of the fitting parameter $\lambda$ for different regimes of $n$. While $\lambda$ is determined by the broadening $\WG$ at $\EF \ll \hbar\wG/2$ (Eq.~(2)), the coupling constant extracted from $\wG$ (Eq.~(1)) is more sensitive to the strength of the non-adiabatic hardening, which occurs for $\EF>\hbar\wG/2$.
This could point to a charge carrier density dependent electron-phonon-coupling constant.
The increase of $\lambda$ with laser power when extracted via $\WG$ (obtained from fitting with Eq.~(2)) can also be contextualized when considering the similarity of the influence of laser power and $n$: With increasing laser power the number of free charge carriers increases due to photo excitation similar to a gate-induced increase in $n$, which may explain why $\lambda$ approaches the value found for $\EF>\hbar\wG/2$ at larger laser powers.

\begin{figure}
   \includegraphics[draft=false,keepaspectratio=true,clip,width=0.9\columnwidth]{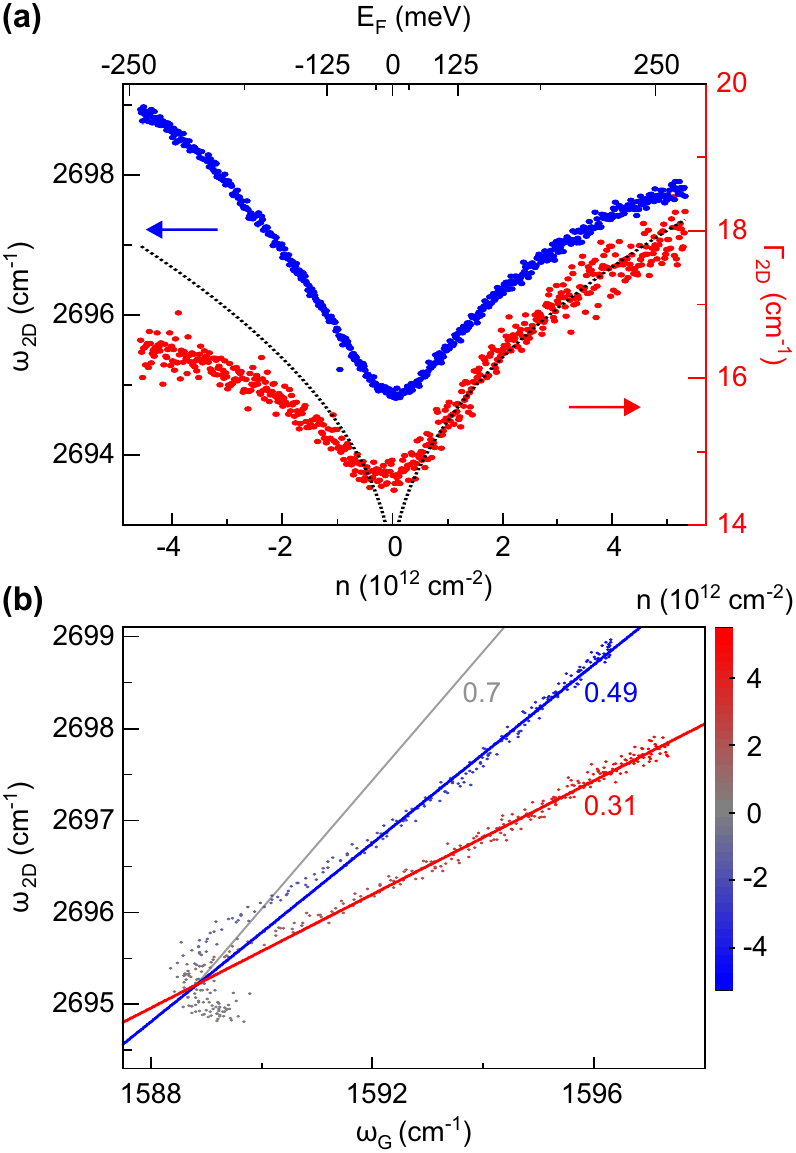}
	\caption{
		(a) Position, i.e. frequency $\wD$ (blue, left axis) and line width $\WD$ (red, right axis) of the Raman 2D~peak as a function of the charge carrier density $n$ at a laser power of $1\,\mathrm{mW}$. The upper axis shows the corresponding Fermi energy $\EF$ assuming $\vF = 0.98\times10^6\,\mathrm{m/s}$. The dotted line shows $\WD$ calculated with Eq.~(\ref{eq:FWHM2DBasko}).
		(b) Scatter plot of $\wD$ as a function of $\wG$ for different charge carrier densities $n$ (see color bar). The colors of the data points correspond to different $n$. The blue (red) line shows a linear fit at high charge carrier density for hole (electron) doping.
		For comparison: The grey line marks a slope of 0.7 (for more details see text).
	}
	\label{fig02}
\end{figure}

\section{Tuning the line width of the 2D peak}

Next, we focus on the charge carrier dependence of the Raman 2D~peak. Even though the precise line shape of the 2D peak is still subject of ongoing research~\cite{Venezuela2011,Neumann2016Dec,Frank2011, Berciaud2013}, it is commonly fitted by a single Lorentzian, especially when used for rapid characterization of graphene and graphene-based heterostructures with regards to doping and strain variations~\cite{Lee2012,Neumann2015b,Banszerus2017}.
As such, we limit the following discussion to a single Lorentzian to provide a useful reference for device characterization.

The extracted positions~$\wD$ and line widths~$\WD$ of the 2D Raman peak are shown in Fig.~\ref{fig02}(a). The frequency $\wD$ shows a significant non-symmetric increase with increasing $|n|$. The increase in $\wD$ can be attributed to a non-adiabatic contribution, similar but weaker to the effect observed in the case of the G~peak~\cite{Das2008Mar,Froehlicher2015May}.
The asymmetry results from a small adiabatic contribution originating from the change in the lattice parameter with doping, which apparently is different for hole and electron doping
~\cite{Das2008Mar}.

Figure \ref{fig02}(b) shows $\wD$ as a function of $\wG$ for different gate-voltage controlled charge carrier densities $n$ (see color bar).
This two-dimensional representation
$(\wG,\wD)$
of the Raman peak positions is commonly used to separate the influence of strain and doping in a kind of "vector decomposition" method~\cite{Lee2012,Banszerus2017}.
This decomposition approach uses the assumption that $\wD(\wG)$ is partially linear.
Evidently, in graphene encapsulated in hBN we find that $\wD(\wG)$ shows a strong non-monotonous contribution at low~$n$ (grey 'tail' in Fig.~\ref{fig02}(b)), due to the appearance of the phonon anomaly of the G-peak (see Fig.~\ref{fig01}(b)).
Only for larger carrier densities $|n|>2\times10^{12}\,\mathrm{cm^{-2}}$ the response linearizes. By means of linear regression in this doping regimes, we find slopes of $\partial\wD/\partial\wG\approx0.49$ and $\partial\wD/\partial\wG\approx0.31$ for hole and electron doping, respectively.
These values differ significantly from the values reported by Lee \etal \cite{Lee2012} ($\partial\wD/\partial\wG\approx0.7$, see gray line in Fig.~3(b)); however they are in agreement with the values found earlier by Froehlicher \etal \cite{Froehlicher2015May} on liquid-gated graphene on SiO\textsubscript{2}.
We stress that special care has to be taken when using this decomposition approach to analyze high quality hBN/graphene/hBN heterostructures, as these samples typically show doping values significantly below $10^{12}\,\mathrm{cm^{-2}}$, where the non-monotonic behavior of $\wG$ and $\wD$ becomes relevant (see Fig.~3(b)).
An increased $\wD$, for example, which results from low doping, could otherwise be misinterpreted as a screening of the Kohn anomaly at the $K$ point also resulting in a stiffening of the 2D-mode phonons~\cite{Banszerus2017,Forster2013Aug}.

Thanks to the high quality of our sample $\WD$ is not dominated by the broadening due to nanometer-scale strain variations~\cite{Neumann2015b} and shows an ultimately low value of $\WD\approx14.5\,\cm$.
This allows us to observe the significant monotonic increase in $\WD$ with $|n|$, see red data points in Fig.~\ref{fig02}(a).
This increase can be attributed to an increased electron-electron scattering rate $\gamma_\mathrm{ee}$ with increasing charge carrier density. Basko~\cite{Basko2008} calculated that the line width of the 2D-peak is given by
\begin{equation}
	\WD=8\sqrt{2^{2/3}-1}\frac{v_\mathrm{TO}}{\vF}\gamma_\mathrm{eh},
	\label{eq:FWHM2DBasko}
\end{equation}
where $\vF$ and $v_\mathrm{TO}$ are the Fermi velocity and the (transverse optical) phonon velocity.
The ratio of $v_\mathrm{TO}$ and $\vF$ is determined by the strength of the Kohn anomaly at the $K$-point and can be extracted by measuring the dispersion of $\wD$ with laser energy. Berciaud~\etal \cite{Berciaud2013} reported a value of $v_\mathrm{TO}/\vF\approx6.2\times10^{-3}$ at a laser energy of $E_\mathrm{L}\approx2.33\,\mathrm{eV}$. The electronic broadening parameter $\gamma_\mathrm{eh}$ increases due to electron-electron scattering $\gamma_\mathrm{ee}$ as shown by Basko \etal \cite{Basko2009}, which calculated that the electron-electron scattering rate scales linearly with $\EF$:
\begin{equation}
	\gamma_\mathrm{ee}\approx 0.06 |\EF|.
	\label{eq:gammeee}
\end{equation}
Note that the exact prefactor depends on the dielectric screening of the electron-electron interaction.
Here, we assume $\epsilon\approx5$, which approximates the mean of the in-plane and out-of-plane dielectric constant of hBN~\cite{Laturia2018Mar}. When considering that the electronic broadening $\gamma_\mathrm{eh}=\gamma_\mathrm{eh}^0+ \gamma_\mathrm{ee}$ increases due electron-electron scattering $\gamma_\mathrm{ee}$ as described in Eq.~(\ref{eq:gammeee}), we find qualitative agreement between Eq.~(\ref{eq:FWHM2DBasko}) and our experimentally extracted $\WD$, see black dotted line in Fig.~\ref{fig02}(a).
Here we use $\gamma_\mathrm{eh}^0\approx43.3\,\mathrm{meV}$, which corresponds to a residual peak width of $\WD^0\approx13.3\,\cm$ due to phonon anharmonicities and intrinsic electronic broadening~\cite{Neumann2015}. Note that the experimental data shows an unexpected asymmetry not captured in Eq.~(\ref{eq:FWHM2DBasko}).

\begin{figure}
	\includegraphics[draft=false,keepaspectratio=true,clip,width=0.95\columnwidth]{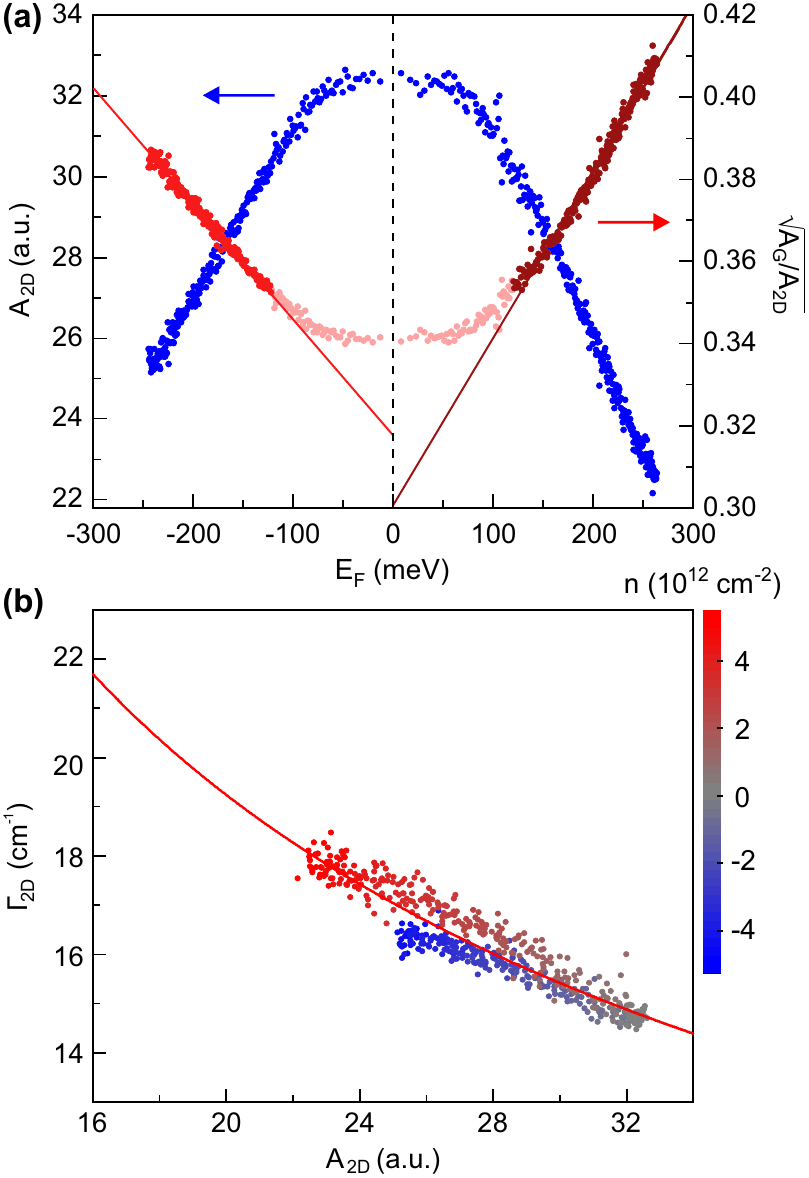}
	\caption{
		(a) Area (blue, left axis) of the 2D~peak as a function of the Fermi energy $\EF$ under the assumption of $\vF = 0.98\times10^6\,\mathrm{m/s}$ at a laser power of $1\,\mathrm{mW}$.
		The right axis shows the normalized root of the area $\sqrt{A_\mathrm{G}/A_\mathrm{2D}}$ in red. The red solid lines are fits to the data points based on Eq.~(6) for both hole doping (red) and electron doping (dark red). The light red data points are omitted for both fits.
		(b) $\WD$ as a function of $\AD$ showing the expected behavior $\WD\propto1/\sqrt{\AD}$ (red curve). See text for details. The colors of the data points correspond to different gate-controlled charge carrier densities~$n$.
		After combing Eqs.~(3) and (5) we used $\WD=a/\sqrt{A_{2D}}+c$ to fit the data in panel (b). The obtained values are $a=2942$ (arbitrary due to $A_{2D}^0$) and $c=-1.57\pm 0.28\,\mathrm{cm^{-1}}$.
	}
	\label{fig03}
\end{figure}

To further elucidate the influence of electron-electron scattering, we now focus on the area of the 2D~peak $\AD$.
The blue data in Fig.~\ref{fig03}(a) shows that $\AD$ decreases with increasing $|\EF|$, which results from the increased electron-electron scattering rate $\gamma_\mathrm{ee}$ with $|n|$. Following Basko \etal \cite{Basko2008,Basko2009} the area
of the 2D peak is given by
\begin{equation}
	A_\mathrm{2D}\propto \left(\frac{\gamma_\mathrm{K}}{\gamma_\mathrm{e\text{-}ph}+\gamma_\mathrm{ee}}\right)^2,
	\label{eq:Intensity2D}
\end{equation}
where the total electron scattering rate $\gamma_\mathrm{e\text{-}ph}+\gamma_\mathrm{ee}$ is given by the sum of the electron-phonon scattering rate $\gamma_\mathrm{e\text{-}ph}$, the electron-electron scattering rate $\gamma_\mathrm{ee}$ and the electron-defect scattering rate has been neglected.
The electron-phonon scattering rate is $\gamma_\mathrm{e\text{-}ph}=\gamma_\mathrm{K}+\gamma_\mathrm{\Gamma}$~\cite{Basko2008,Basko2009}, where $\gamma_\mathrm{K}$ is the scattering rate of the optical phonon at $K$ and $\gamma_\mathrm{\Gamma}$ takes into account the scattering from optical phonons at the $\Gamma$ point. While $\gamma_\mathrm{K}$ and $\gamma_\mathrm{e\text{-}ph}$ should not depend on the charge carrier density, $\gamma_\mathrm{ee}$ scales with $\EF$ (see Eq.~(\ref{eq:gammeee})).
As the area of the G-peak, $A_\mathrm{G}$, does not depend on $\EF$ for experimentally viable $|\EF|<E_\mathrm{L}/2$, it is useful to normalize $A_\mathrm{2D}$ to $A_\mathrm{G}$. Eq.~(\ref{eq:Intensity2D}) can then be transformed into:
\begin{equation}
	\sqrt{\frac{A_\mathrm{G}}{A_\mathrm{2D}}}=\sqrt{\frac{A_\mathrm{G}^0}{A_\mathrm{2D}^0}}\cdot \left(1+\frac{0.06}{\gamma_\mathrm{e\text{-}ph}}|E_\mathrm{F}|\right),
	\label{eq:SqrtIntensity2D}
\end{equation}
where $A_\mathrm{G/2D}^0$ denote the respective areas at $\EF=0$. The red data in Fig.~\ref{fig03}(a) depict the experimentally obtained $\sqrt{A_\mathrm{G}/A_\mathrm{2D}}$ values. Akin to the asymmetry found in $\WD$, we find that $\sqrt{A_\mathrm{G}/A_\mathrm{2D}}$ differs between electron and hole doping.
By fitting Eq.~(\ref{eq:SqrtIntensity2D}) separately to the electron and hole regime, we find electron-phonon scattering rates of $\gamma_\mathrm{e\text{-}ph}^e=(67.5\pm0.8)\,\mathrm{meV}$ and $\gamma_\mathrm{e\text{-}ph}^h=(44.2\pm0.5)\,\mathrm{meV}$, respectively.
While previous measurements on electrochemically gated graphene on SiO\textsubscript{2} found scattering rates of similar magnitude (39-72~meV), no asymmetry between electron and hole doping was reported~\cite{Froehlicher2015May}.

The extracted scattering rate is directly connected to the dimensionless electron-phonon coupling constants $\lambda_\Gamma$ and $\lambda_{K}$, which describe the coupling strength of the optical phonons at the $\Gamma$ (G-mode) and near the $K$ point (2D-mode), respectively. Following Refs.~\cite{Basko2008,Basko2009,Basko2008Jan,Froehlicher2015May} we obtain
\begin{equation}
	\gamma_\mathrm{e\text{-}ph}=\gamma_\mathrm{K}+\gamma_\mathrm{\Gamma}=\frac{\pi\lambda_{K}}{2}\left(\frac{E_\mathrm{L}}{2}-E_{D}\right)+\frac{\pi\lambda_\mathrm{\Gamma}}{2}\left(\frac{E_\mathrm{L}}{2}-E_\mathrm{G}\right),
	\label{eq:LambdaK}
\end{equation}
where $E_\mathrm{G}\approx197\,\mathrm{meV}$ and $E_\mathrm{D}\approx167\,\mathrm{meV}$ are the optical phonon energies at $\Gamma$ and near the $K$ point, respectively.
By using $\lambda_\mathrm{\Gamma}\approx4.4\times10^{-3}$ as extracted from Eqs.~(\ref{eq:PhononAnomaly}) and~(\ref{eq:PhononAnomalyGamma}), we find in our high quality hBN/graphene/hBN sample that the electron-phonon coupling strength near the $K$ point is $\lambda_{K}\approx 37.9\times10^{-3}$ and $\lambda_{K}\approx 23.4\times10^{-3}$ for electron and hole doping, respectively. These values are in agreement~\cite{com1}
with previous reports measured on electrochemically gated graphene on SiO\textsubscript{2} ($\lambda_{K}\approx 27\times10^{-3} - 39\times10^{-3}$)~\cite{Froehlicher2015May}.

Finally, we highlight the common origin of the decrease of $\AD$ and of the increase in $\WD$ with increasing $|n|$ by plotting $\WD$ as a function of $\AD$, see Fig.~\ref{fig03}(b). Indeed, there is an universal scaling of the 2D peak width with the area $\WD\propto1/\sqrt{\AD}$. This functional connection can easily be computed by combining Eqs.~(\ref{eq:FWHM2DBasko}) and~(\ref{eq:Intensity2D}). While this connection was predicted theoretically~\cite{Neumann2015b,Venezuela2011}, it is finally possible to verify this scaling experimentally, thanks to the negligible amounts of other 2D line broadening effects in our sample.

\section{Conclusion}

In summary, we thoroughly investigated the Raman spectra of high-quality, i.e. ultra-flat graphene encapsulated in hBN, especially with regards to its charge carrier density dependence.
In the first part of this work we focused on the electron coupling to the G-mode phonon. We showed a well visible phonon anomaly of the G-mode in graphene and discussed its laser power dependence.
Interestingly, we observe indications that the electron-phonon coupling $\lambda$ might significantly depend on the charge carrier density.
Taking the 2D-peak into account we provide a new benchmark for the analysis of the Raman spectra of hBN/graphene/hBN heterostructures in regards to the charge carrier density via the so-called "vector decomposition" method. Furthermore, we extract the electron-phonon coupling strength at $K$ by examining the electron-electron scattering induced drop in the 2D-peak intensity and show how the electron-electron scattering leads to a broadening of the 2D~peak. We believe that this systematic study provides an unprecedented reference for Raman spectroscopy on high-quality graphene samples encapsulated in hBN and is useful to further investigations on the electron-phonon coupling and to benchmark graphene samples (including graphene-based heterostructures and twisted bilayer graphene) also in the context of using Raman spectroscopy for process monitoring.

\begin{acknowledgments}

\textbf{Acknowledgments:}
The authors thank F.~Mauri, C.~Neumann and S.~Reichardt for helpful discussions.
This project has received funding from the European Union’s Horizon 2020 research and innovation programme under grant agreement No. 881603 (Graphene Flagship) and from the European Research Council (ERC) under grant agreement No. 820254, under FLAG-ERA grant TATTOOS and 2D-NEMS by the Deutsche Forschungsgemeinschaft (DFG, German Research Foundation) - 437214324 and 436607160, the Deutsche Forschungsgemeinschaft (DFG, German Research Foundation) under Germany’s Excellence Strategy - Cluster of Excellence Matter and Light for Quantum Computing (ML4Q) EXC 2004/1 - 390534769, through DFG (STA 1146/11-1), and by the Helmholtz Nano Facility~\cite{Albrecht2017}. Growth of hexagonal boron nitride crystals was supported by JSPS KAKENHI (Grant Numbers 19H05790, 20H00354 and 21H05233).

\end{acknowledgments}

\section*{Appendix A: Transport Measurements}

\begin{figure}[b!]
    \includegraphics[draft=false,keepaspectratio=true,clip,width=0.95\columnwidth]{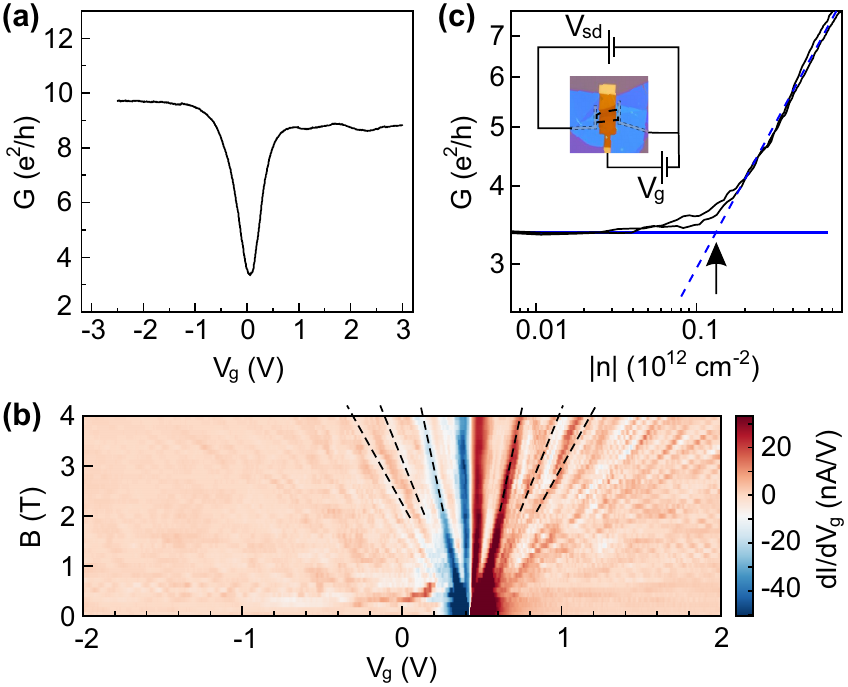}
	\caption{
(a) Two-terminal electrical conductance $G$ as a function of gate voltage $V_\mathrm{g}$ of the locally gated hBN/graphene/hBN heterostructure after laser illumination.
(b) Two-terminal Landau fan measurement (recorded before laser illumination) showing the differential current $dI/dV_\mathrm{g}$ as function of applied gate voltage $V_\mathrm{g}$ and magnetic field ($B$). The bias voltage is $V_\mathrm{sd}=100\,\mathrm{\mu V}$. By analysing the slopes of the observed Shubnikov-de Haas oscillations (see dashed lines) we can extract the lever arm. For more details see text and Ref.~\cite{Sonntag2018May}.
(c) Double logarithmic graph of the  conductance $G$ after illumination. The dashed and solid lines are linear fits. The crossing point of these lines defines $n^*$ (see arrow). The inset shows the wiring of the sample.
	}
	\label{fig05}
\end{figure}
In Fig.~5 we show low-temperature (4.2~K) transport measurements on the very same graphene sample where the detailed gate-dependent Raman spectroscopy measurements have been performed. Fig.~5(a) shows two-terminal conductance $G$ measurements (source and drain contacts are highlighted in the inset of Fig.~5(c)) as a function of gate voltage $V_g$ after the graphene sample has been illuminated with the green Raman laser.
From the data we extract the charge neutrality point to be at $V_g^0=37$~mV.

A two-terminal Landau fan measured before the illumination with the laser is shown in Fig.~5(b). This plot shows the differential conductance $dI/dV_g$ as function of $V_g$ and applied out-of-plane magnetic field ($B$). The observed Shubnikov-de Haas oscillations in the so-called Landau fan are used to extract the gate lever arm $\alpha= 1.8 \times 10^{12}$ 1/(V cm$^2$), which agrees well with the capacitor model leading to $\alpha=\varepsilon_0 \varepsilon_r /(e d)$, where $\varepsilon=3.4$ is the out-of-plane dielectric constant of hBN~\cite{Pierret2022Jun} and $d \approx 10$~nm is the thickness of the bottom hBN crystal (see schematic in Fig.~1(b)). For more details on this technique please see Ref.~\cite{Sonntag2018May}.

Figure~5(c) shows a double logarithmic graph of the conductance $G$ as function of carrier density $n$. Following Ref.~\cite{Couto2014} we extract from the intersection of the linear fits (see dashed and solid blue lines) the residual charge carrier density inhomogeneity $n^* \approx 13 \times 10^{10}$~cm$^{-2}$. Interestingly, this value matches well with the charge carrier density disorder induced effective temperature $T^*_\mathrm{eff}$ as discussed in the main text.

\section*{Appendix B: Laser power dependence of additional fitting parameters}

\begin{figure}
    \includegraphics[draft=false,keepaspectratio=true,clip,width=1.0\columnwidth]{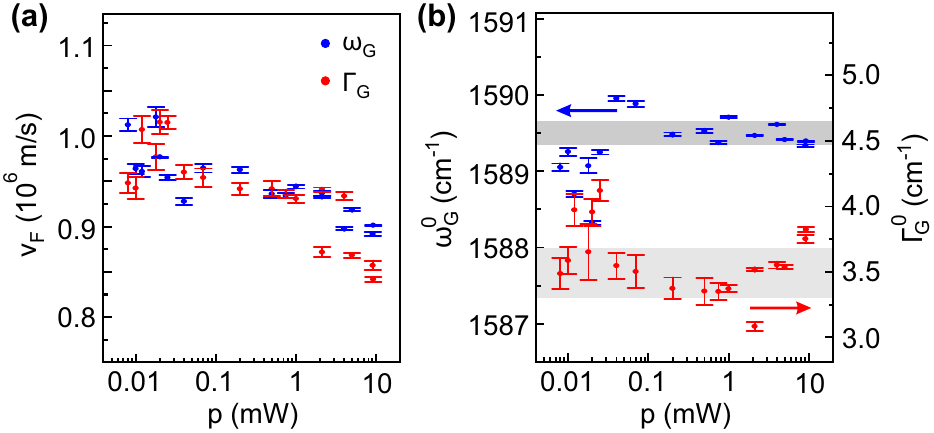}
	\caption{
	(a) Extracted Fermi velocity $\vF$ as a function of laser power $p$; from fitting both $\wG(n)$ (blue data points) and $\WG(n)$ (red data points) with Eq.~(\ref{eq:PhononAnomaly}) and Eq.~(\ref{eq:PhononAnomalyGamma}), respectively.
	(b)~Extracted $\wG^0$ (blue data points) and $\WG^0$ (red data points) as a function of laser power $p$; again from fitting $\wG(n)$ (blue) and $\WG(n)$ (red) with Eq.~(\ref{eq:PhononAnomaly}) and Eq.~(\ref{eq:PhononAnomalyGamma}), respectively.
}
	\label{fig06}
\end{figure}

In Figs.~6(a) and 6(b) we show -- in complete analogy to Figs.~2(c) and 2(d) -- the additional laser power dependent fitting parameters obtained from fitting (i) Eq.~(1) to the data presented in Fig.~2(a) (blue data points) and (ii) Eq.~(2) to data as shown in Fig.~2(b) (red data points).
From both fits we find that there is a consistent $v_\mathrm{F}$, which is only weakly depending on the laser power $p$ (Fig.~6(a)).
Please Note that $\vF$ can also vary with $n$~\cite{Elias2011,Yu2013,Sonntag2018May}. However, as the variation in $\vF$ is mostly located at the CNP and is not pronounced in hBN/graphene/hBN heterostructures, we neglect the $\vF$ renormalization in this work.

Most importantly, we observe (as shown in Fig.~6(b)) that $\wG^0$ (blue data) and $\WG^0$ (red data) are indeed nearly constant as function of laser power $p$ as mentioned in the main text.

\end{document}